\documentclass{article}
\usepackage{spconf,amsmath,graphicx,hyperref,cite}

\title{ReCQR: Incorporating conversational query rewriting to improve Multimodal Image Retrieval}
\name{Yuan Hu \qquad ZhiYu Cao \qquad PeiFeng Li  \qquad QiaoMing Zhu}
\address{School of Computer Science and Technology, Soochow University, Suzhou, China}
\begin{document}
 \ninept
\maketitle

\begin{abstract}
With the rise of multimodal learning, image retrieval plays a crucial role in connecting visual information with natural language queries. Existing image retrievers struggle with processing long texts and handling unclear user expressions. To address these issues, we introduce the conversational query rewriting (CQR) task into the image retrieval domain and construct a dedicated multi-turn dialogue query rewriting dataset. Built on full dialogue histories, CQR rewrites users’ final queries into concise, semantically complete ones that are better suited for retrieval. Specifically, We first leverage Large Language Models (LLMs) to generate rewritten candidates at scale and employ an LLM‑as‑Judge mechanism combined with manual review to curate approximately 7,000 high‑quality multimodal dialogues, forming the ReCQR dataset. Then We benchmark several SOTA multimodal models on the ReCQR dataset to assess their performance on image retrieval. Experimental results demonstrate that CQR not only significantly enhances the accuracy of traditional image retrieval models, but also provides new directions and insights for modeling user queries in multimodal systems.
\end{abstract}

\section{Introduction}
\label{sec:intro}
\begin{figure}[t] 
\centering 
\includegraphics[width=\linewidth,height=0.3\textheight,keepaspectratio]{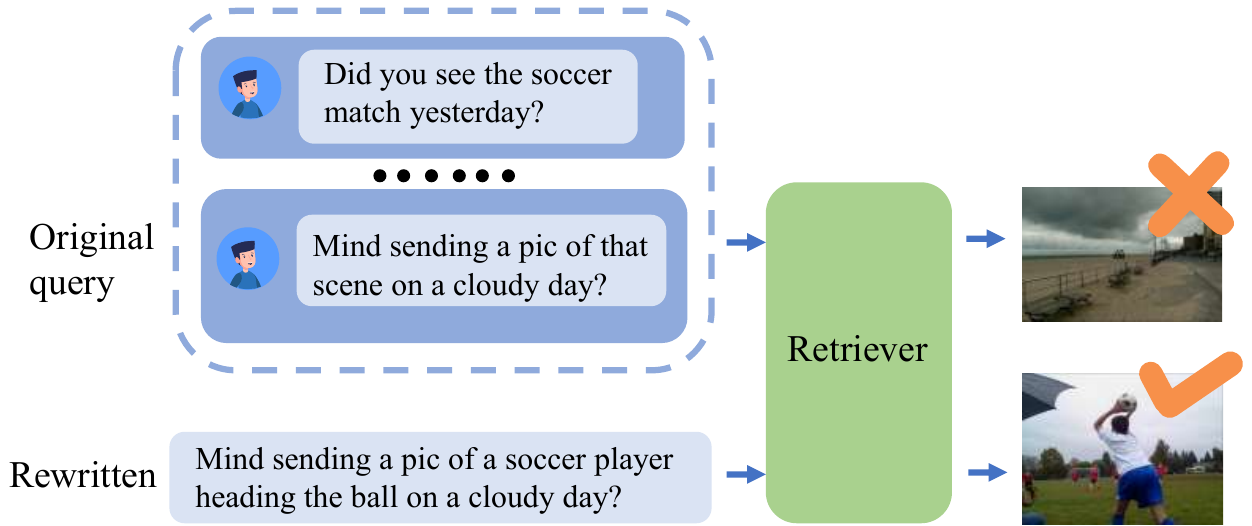} 
\caption{The necessity of conversational query rewriting. The Original Query of user contains an ambiguous reference ("that scene"), which leads to incorrect retrieval results. The Rewritten Query disambiguates the request by incorporating key visual context from the dialogue history, enabling the retriever to return the correct image.} 
\label{fig:introduction}
\end{figure}

Multimodal image retrieval aims to accurately locate target images by comprehending user intentions across both visual and textual contexts. This task becomes particularly crucial in conversational settings, where understanding the dialogue history is essential for interpreting user queries. However, the inherent ambiguity of user requests presents a major challenge for multimodal image retrieval. As shown in Fig.~\ref{fig:introduction}, the final queries are often context-dependent and semantically incomplete, laden with references (e.g.,``that scene on a cloudy day'') that are irresolvable without the dialogue history. While models like CLIP~\cite{radford2021clip} excel at single-turn retrieval, they struggle in conversational settings, due to an inability to resolve these references in dialogue.

Recent conversational image retrieval (CIR) methods~\cite{yuan2021fashion},\cite{zhao2025chatsearch} attempt to address this issue by encoding the entire dialogue history alongside the current query. However, they often introduce noise and redundancy, complicating the retrieval process. An alternative strategy, proven highly effective in text-only domains, is Conversational Query Rewriting (CQR)~\cite{wu2022conqrr, qian2022explicit, ye2023infocqr}, which converts context-dependent utterances into self-contained, retrieval-friendly queries. 
Although initial efforts like McQueen~\cite{yuan2022mcqueen} and JDDC~\cite{zhao2022jddc} have explored multimodal CQR, they focus on visual grounding or general dialogue understanding, not the performance of image retrieval. This leaves a critical gap: the potential of CQR to bridge advanced but static vision-language models with dynamic, multimodal conversations remains largely unexplored.

To bridge this gap, we argue for the timely integration of CQR into multimodal image retrieval. By rewriting the final query into a concise, intent-rich representation (e.g., transforming the query in Fig.~\ref{fig:introduction} into ``Mind sending a pic of a soccer player heading the ball on a cloudy day''), CQR offers a direct pathway to leverage powerful off-the-shelf retrievers (e.g., CLIP) in multi-turn settings.

\begin{figure*}[t] 
\centering 
\includegraphics[width=\linewidth,height=0.4\textheight,keepaspectratio]{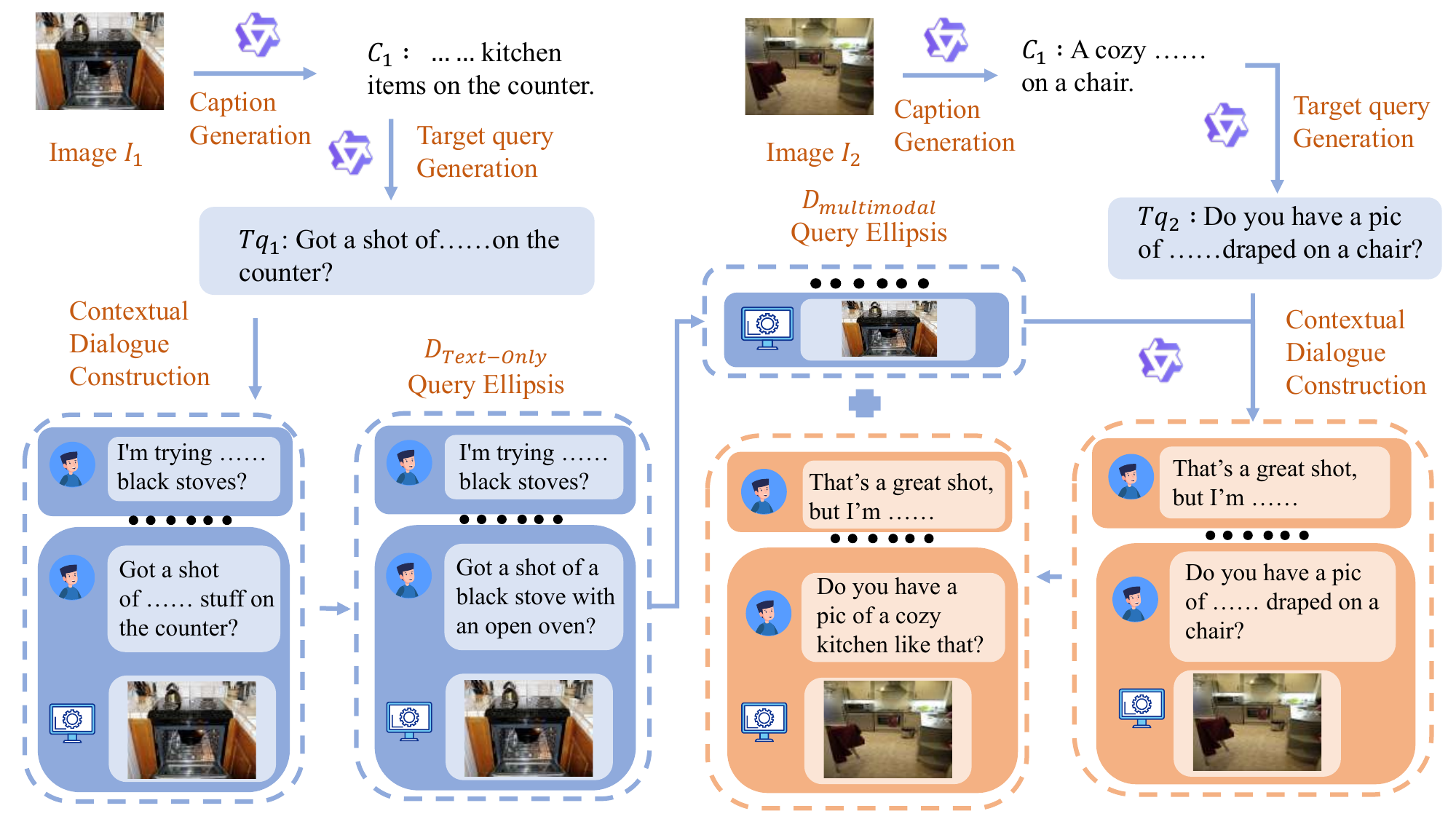} 
\caption{The two-stage dataset construction pipeline. Stage One creates dialogues for a single image $I_1$, generating a standard caption $C_1$, a target query $Tq_1$, a dialogue history $D_1$ and a contextually abbreviated original query $Oq_1$. Stage Two creates dialogues for a semantically related image pair ($I_1$, $I_2$), generating captions ($C_1$, $C_2$), a dialogue $D_2$ that bridges both images, and a final original query $Oq_2$ that resolves to target query $Tq_2$.} 
\label{fig:dataconstruct}
\end{figure*}
We introduce ReCQR, the first benchmark for Conversational Query Rewriting in image retrieval. ReCQR contains~7K multi-turn dialogues with high-quality rewriting, built using a scalable pipeline with LLM generation and LLM-as-Judge filtering, followed by human verification. Extensive benchmarking shows that query rewriting substantially improves multimodal model performance, confirming its value as a core component in future multimodal dialogue systems. Our contributions are threefold:~(1)~We extend CQR to multimodal image retrieval;~(2)~We construct the ReCQR dataset for multimodal CQR evaluation;~(3)~We establish a comprehensive benchmark demonstrating that query rewriting effectively enables off-the-shelf retrieval models to handle complex multimodal dialogues.

\section{Dataset Construction}
\label{sec: Dataset Construction}
Our dataset is constructed from the MSCOCO corpus~\cite{lin2014mscoco} using a two-stage pipeline designed to support conversational query rewriting for multimodal image retrieval. In Stage One, we generate text-only dialogues containing context-dependent queries requiring history resolution, simulating elliptical references from real conversations. Stage Two introduces multimodal dialogues where users reference multiple images, creating cross-modal dependencies. This approach provides training data for models to transform ambiguous queries into self-contained representations suitable for off-the-shelf retrievers like CLIP.
\subsection{Stage One: Text-Only Dialogues Construction}
\label{Single-Image}
The construction pipeline for single-image dialogues is illustrated in Fig.~\ref{fig:dataconstruct}. We begin by sampling 6,000 images from MSCOCO. For each image $I_1$, we generate a standard caption $C_1$ using Qwen2.5-VL-7B-Instruct \cite{bai2025qwen25vltechnicalreport} with the prompt template $P{\text{cap}}$ (see Table~\ref{tab:Prompt}):

\begin{align}
C_1 &= \mathcal{LLM}{\text{cap}}(I_1; P{\text{cap}}),
\end{align}

Subsequently, the Qwen3-14B model \cite{yang2025qwen3technicalreport} performs the following three sequential operations:

(1) \textbf{Target Query Generation:} This step aims to generate a clear and self-contained retrieval target from the image caption, serving as the ground truth for the conversational query.
\begin{align}
Tq_1 &= \mathcal{LLM}{\text{tar}}(C_1; P{\text{tar}}),
\end{align}
where $\mathcal{LLM}{\text{tar}}$ denotes the large language model generating the target query $Tq_1$ from image caption $C_1$ using prompt $P{\text{tar}}$. This produces a complete, standalone expression of user retrieval intent.

(2) \textbf{Contextual Dialogue Construction:} This step constructs a realistic multi-turn dialogue history based on the target query.
\begin{align}
D_1 &= \mathcal{LLM}{\text{dia}}(Tq_1; P{\text{dia}}),
\end{align}
where $\mathcal{LLM}{\text{dia}}$ denotes the large language model constructing a multi-turn dialogue history $D_1$ from target query $Tq_1$ using dialogue-specific prompt $P{\text{dia}}$.

(3) \textbf{Query Ellipsis:} This step produces an elliptical or context-dependent user query by removing information that can be inferred from the dialogue history.
\begin{align}
E_D &= \mathcal{E}(D_1),~E_T = \mathcal{E}(Tq_1), ~E{\text{overlap}}= E_D \cap E_T, \\
Oq_1 &= \mathcal{LLM}{\text{ellip}}(D_1, Tq_1, E{\text{overlap}}; P{\text{ellip}}, \mathcal{S}),
\end{align}
where $\mathcal{E}$ denotes the spaCy~\cite{spacy},~\cite{en_core_web_sm} entity extractor; $E_D$ and $E_T$ represent the entity sets extracted from dialogue history $D_1$ and target query $Tq_1$, respectively; $E{\text{overlap}}$ represents the intersection of entities; $P{\text{ellip}}$ is the ellipsis generation prompt; and $\mathcal{LLM}{\text{ellip}}$ generates the original query $Oq_1$ by obfuscating information inferable from $D_1$, using few-shot examples $\mathcal{S}$ sampled from a predefined set. This process effectively simulates conversational phenomena requiring deep contextual understanding.

The final constructed data instance follows the format:~$\mathcal{D}_{\text{Text-Only}} = D_1 \oplus Oq_1 \oplus I_1$, where $D_1 = {u_1, s_1, \dots, u_n, s_n}$ represents the multi-turn dialogue history. Each $u_i$ and $s_i$ contains exclusively textual content, denoting user utterances and system responses, respectively. $Oq_1$ denotes the original query serving as model input, which is a more colloquial and context-dependent formulation compared to $Tq_1$, simulating natural conversational expressions typical in daily communication. $I_1$ represents the target image for retrieval. The operator $\oplus$ indicates the concatenation of these components into a complete data instance.

This structure ensures that the dialogue context contains exclusively textual exchanges while maintaining the association between the original query $Oq_1$ and its corresponding target image $I_1$.

\begin{table}[t]
\centering
\caption{Prompt Templates Used in Dataset Construction}
\label{tab:Prompt}
\begin{tabular}{lcccc}
\hline
\textbf{Content}& \multicolumn{4}{c}{\textbf{Context}}\\
\hline
$ P{\text{cap}}$& \multicolumn{4}{c}{Briefly describe this image in one sentence ...}\\
$ P{\text{tar}}$& \multicolumn{4}{c}{You are .. a picture that  looks like this one.}\\
$ P{\text{dia}}$& \multicolumn{4}{c}{You are a helpful vision-language assistant …}\\
$ P{\text{dia2}}$& \multicolumn{4}{c}{You are continuing a conversation ......}\\
$ P{\text{elip}}$& \multicolumn{4}{c}{Your ... rewrite the user's final image request….}\\
 \hline
\end{tabular}
\end{table}
\subsection{Stage Two: Multimodal Dialogues Construction}
\label{Multi-Image}
Fig.~\ref{fig:dataconstruct} details the construction of Multimodal dialogues. From MSCOCO, we sampled 12,000 additional images (distinct from Stage One). These images were then paired with those from the first stage to form candidate pairs $(I_1, I_2)$. For each first-stage image, multiple semantically related candidates could be identified, in such cases, we randomly retained one pair to ensure diversity. This second stage introduces mixed-modality contexts, simulating real-world scenarios where users may reference both dialogue history and previously shared images.

\begin{figure}[t] 
\centering 
\includegraphics[width=\linewidth,height=0.4\textheight,keepaspectratio]{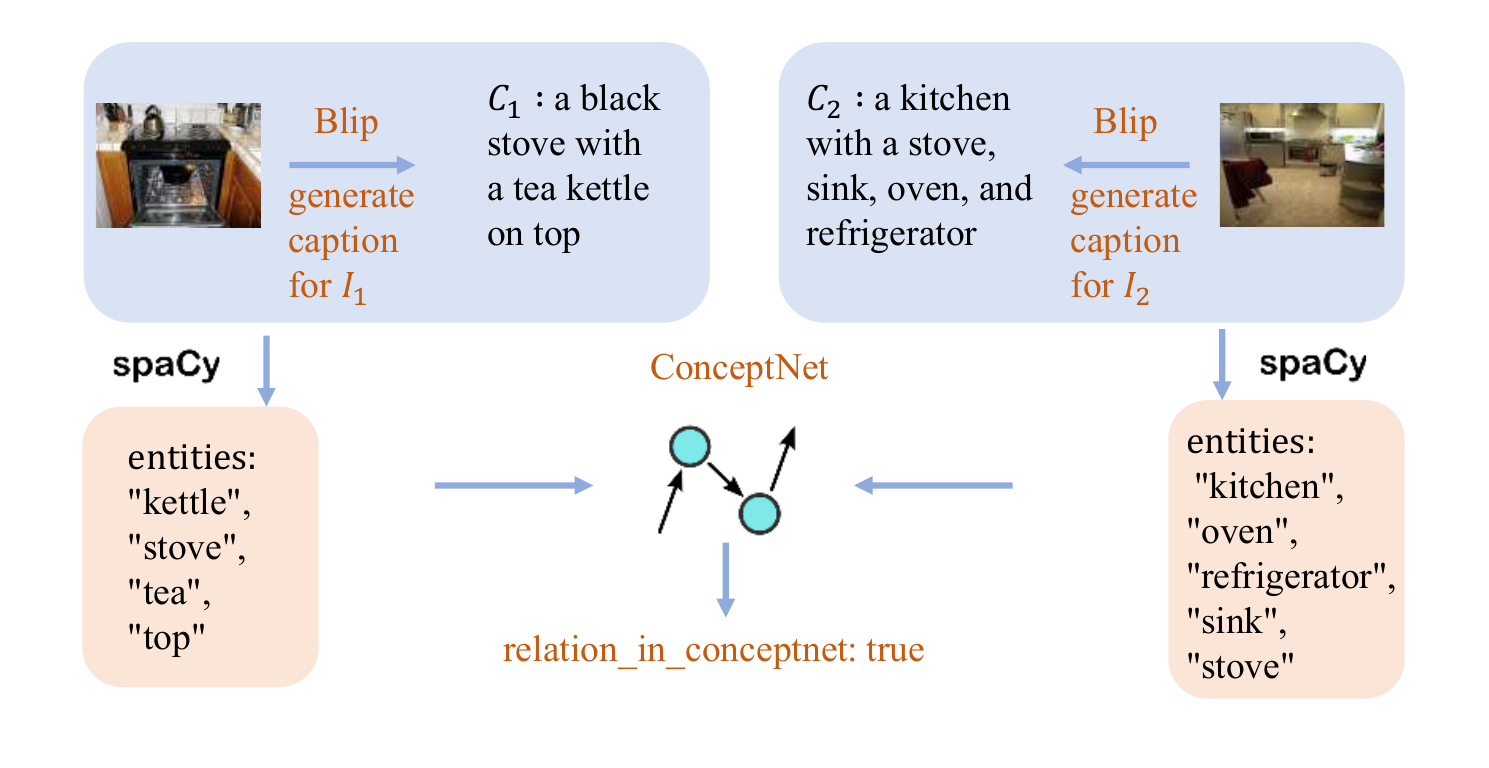} 
\caption{Semantic relevance validation for image pairing. Candidate images $I_1$ and $I_2$ are processed by BLIP to generate captions. Key entities are extracted from these captions (using spaCy) and their relationship is checked in the ConceptNet knowledge graph.} 
\label{fig:concept}
\end{figure}

The process for each validated pair involves two main steps:

(1) \textbf{Semantic Pairing:} To ensure semantic relevance and topical consistency (Fig.~\ref{fig:concept}), we employed structured semantic verification.
\begin{align}
C_1, C_2 &= \text{BLIP}(I_1), \text{BLIP}(I_2), \\
E_1, E_2 &= \mathcal{E}(C_1), \mathcal{E}(C_2), \\
\text{Pair}(I_1, I_2) & \leftarrow \exists r \in \text{ConceptNet}(E_1, E_2),
\end{align}
where BLIP \cite{li2022blip} denotes the caption generation model, $\mathcal{E}$ represents the spaCy entity extractor; $E_1$ and $E_2$ are the sets of semantic entities extracted from captions $C_1$ and $C_2$, respectively; and ConceptNet \cite{speer2018conceptnet55openmultilingual} verifies the existence of any semantic or commonsense relation between the extracted entities. Only image pairs with confirmed relationships were retained for subsequent dialogue construction.

(2) \textbf{Dialogue and Query Generation:} The construction follows a procedure similar to Stage One with enhanced contextual inputs.
\begin{align}
C_1, C_2 &= \mathcal{LLM}{\text{cap}}(I_1; P{\text{cap}}),~\mathcal{LLM}{\text{cap}}(I_2; P{\text{cap}}), \\
Tq_2 &= \mathcal{LLM}{\text{tar}}(C_2; P{\text{tar}}), \\
D_2 &= D_{\text{Text-Only}} \oplus\mathcal{LLM}{\text{dia}}(Tq_2, D_1, C_1; P{\text{dia2}}), \\
E_D &= \mathcal{E}(D_2),~E_T = \mathcal{E}(Tq_2), ~E{\text{overlap}}= E_D \cap E_T, \\
Oq_2 &= \mathcal{LLM}{\text{ellip}}(D_2, Tq_2, E{\text{overlap}}; P{\text{ellip}}, \mathcal{S}),
\end{align}
where $\mathcal{LLM}{\text{dia}}$ generates the multi-turn dialogue $D_2$ by extending the existing dialogue history $D_1$ from Stage One, incorporating both the target query $Tq_2$ and the caption $C_1$ of the first image to ensure contextual continuity and semantic coherence across images.

The final data instance follows the format:
$\mathcal{D_{\text{multimodal}}} = D_2 \oplus Oq_2 \oplus I_2$,
where $D_2$ contains the complete multi-image dialogue history, $Oq_2$ denotes the original query serving as model input, and $I_2$ serves as the retrieval target.

This two-stage approach produces a dataset spanning simple to complex contexts, enabling development of models for context-aware multimodal query rewriting.

\subsection{Dataset Quality Control}
To ensure high quality, we employed a rigorous two-stage filtering process. First, every (dialogue, original query, target query) triplet was automatically evaluated by GPT-4~\cite{achiam2023gpt}. Each triplet was scored on a 5-point scale based on contextual coherence, appropriate information omission, and query reconstructability. Triplets receiving a score below 3 were automatically discarded. The remaining triplets then underwent manual review. Each triplet was independently evaluated by two human annotators using the same criteria as the automated evaluation, but making binary judgments (accept/reject) instead of numerical scores. To resolve annotation conflicts, a third expert annotator was introduced as a tie-breaker, with their decision being final. This multi-stage, consensus-driven approach guarantees that the final dataset maintains both natural conversational flow and precise grounding in the visual context. Following this rigorous filtering process, we retained a high-quality dataset of 4,000 for the single-image portion and 3,000 for the multi-image portion.~The detailed data split is presented in Table~\ref{tab:dataset_split}.

\section{Task and Evaluation}
\label{sec:Task and Evaluation}
\subsection{Task Formulation}
The core task of Retrieval-Oriented Conversational Query Rewriting (ReCQR) extends conversational query rewriting to the multimodal domain for text-to-image retrieval. Given a multimodal dialogue $D = {u_1, s_1, \dots, u_{n}, s_{n}}$ (containing user utterances $u_i$ and system responses $s_i$) and the current user query $Oq$, the goal is to generate a rewritten query $\hat{q}$ that resolves contextual references (e.g., coreference and ellipsis), incorporates relevant visual information, and forms an explicit, self-contained query suitable for retrieval. This is formalized as: $\hat{q} = \mathcal{F}(H, Oq)$ is evaluated via off-the-shelf retrieval performance, where $\mathcal{F}$ denotes the query rewriting model.

\subsection{Evaluation Metrics}
\label{sec:Evaluation Metrics}
To objectively evaluate the quality of the rewritten queries, we adopt Recall@K (R@K) as our primary evaluation metric following the standard information retrieval protocol. We specifically report R@1, R@5, and R@10 to comprehensively assess the retrieval accuracy across different levels of result granularity, thereby providing a multi-faceted perspective on model performance.

\subsection{Baselines}
\label{ssec:Baseline Models and Experimental Setup}

Our experimental framework consists of Conversational Query Rewriting (CQR) models and a fixed image retrieval backbone.~We evaluate three multimodal large language models (MLLMs) for the rewriting task:~\textbf{Qwen2.5-VL-7B-Instruct}~\cite{bai2025qwen25vltechnicalreport}, \textbf{LLaVA-v1.6-Mistral-7B-HF}~\cite{liu2024llavanext}, and \textbf{GLM-4.1V-9B-Thinking}~\cite{vteam2025glm45vglm41vthinking}.

For retrieval, we use \textbf{CLIP-ViT-B/32} \cite{radford2021clip} as our fixed backbone model. It encodes both rewritten queries and candidate images into a shared embedding space where retrieval is performed by cosine similarity. Keeping this component constant ensures that any performance variation (R@K) is solely attributable to the quality of the rewritten queries from different CQR models.

\begin{table}[t]
\centering
\caption{Dataset Partitioning}
\label{tab:dataset_split}
\begin{tabular}{lcccc}
\hline
\textbf{Dataset} & \textbf{Train} & \textbf{Validation} & \textbf{Test} & \textbf{Total} \\
\hline
Text-Only& 3,000 & 500 & 500 & 4,000 \\
Multimodal& 2,000 & 500 & 500 & 3,000 \\
\hline
\textbf{Overall Total} & \textbf{5,000} & \textbf{1,000} & \textbf{1,000} & \textbf{7,000} \\
\hline
\end{tabular}
\end{table}

\subsection{Experimental Settings}
We design a two-phase training and evaluation pipeline to rigorously assess model performance and the impact of multimodal context:~(1)~\textbf{Text-Only (T) Setting.} In this initial phase, LLM models are fine-tuned and evaluated using only the \textit{textual} dialogue history. The associated images from the training set are used solely for providing the target query that guides the rewriting process during training, but are not input to the model.  This setting tests the models' ability to perform query rewriting based on textual context alone. The rewritten queries from the text-only test set are then evaluated using the CLIP retrieval backbone;~(2)~\textbf{Multimodal (M) Setting.} This phase builds upon the (T).  Models are initialized with the weights from the text-only phase and are then further fine-tuned on the multimodal dataset. In this setting, the models receive both the textual dialogue history and the images from the history as input. This allows them to leverage visual information to resolve references and enrich the rewrite. The models are subsequently evaluated on the multimodal test set, and their output queries are again used for retrieval with CLIP.

This progressive setup allows us to isolate the contribution of visual grounding to the query rewriting task. We also include key reference points in Table~\ref{tab:grouped_results}: \textbf{Original Query} (lower bound), \textbf{Target Query} (oracle rewrite), and \textbf{Caption} (direct description, performance ceiling).

\subsection{Analysis of Experimental Results}
\label{ssec:Analysis of Experimental Results}

The experimental results presented in Table~\ref{tab:grouped_results} provide a evaluation of the baseline models on the ReCQR task, highlighting the impact of multimodal context and the inherent challenge of the task.

\begin{table}[t]
\centering
\caption{Retrieval performance of rewritten queries on the ReCQR benchmark. All results are obtained using CLIP-ViT-B/32 as the retrieval backbone.}
\begin{tabular}{l|ccc}
\hline
\textbf{Method} & \textbf{R@1} & \textbf{R@5} & \textbf{R@10} \\
\hline
\multicolumn{4}{c}{\textbf{Text-Only Dataset}} \\
\hline
Original query& 3.6 & 10.4 & 13.6 \\
Target query  & 22.4 & 43.6 & 51.2 \\
Caption & \textbf{27.8} & \textbf{49.6} & \textbf{56.4} \\
\hline
Qwen2.5-VL-7B-Instruct   & 13.6 & 25.0 & 31.4 \\
Qwen2.5-VL-7B-Instruct(T) & 19.2 & 37.8 & 44.6 \\
Qwen2.5-VL-7B-Instruct(M) & 18.0 & 37.6 & 45.2 \\
LLaVA-v1.6-Mistral-7B-HF& 11.4 & 26.0 & 31.0 \\
LLaVA-v1.6-Mistral-7B-HF(T)& 18.8 & 38.0 & 44.4 \\
LLaVA-v1.6-Mistral-7B-HF(M)& 17.6 & 36.8 & 43.0 \\
GLM-4.1V-9B-Thinking   & 15.4 & 31.6 & 38.0 \\
GLM-4.1V-9B-Thinking(T) & \textbf{19.6} & \textbf{38.4} & \textbf{45.4} \\
GLM-4.1V-9B-Thinking(M) & 17.6 & 36.4 & 42.4 \\
\hline
\multicolumn{4}{c}{\textbf{Multimodal Dataset}} \\
\hline
Original query & 3.2 & 7.6 & 9.6 \\
Target query   & 20.4 & 38.0 & 48.4 \\
Caption & \textbf{24.2} & \textbf{42.4} & \textbf{52.2} \\
\hline
Qwen2.5-VL-7B-Instruct   & 0.2 & 0.4 & 0.4 \\
Qwen2.5-VL-7B-Instruct(T) & 6.2 & 13.4 & 17 \\
Qwen2.5-VL-7B-Instruct(M) & 11.6 & 24.4 & 30.2 \\
LLaVA-v1.6-Mistral-7B-HF & 2.8 & 8.8 & 12.6 \\
LLaVA-v1.6-Mistral-7B-HF(T)& 7.6 & 16.6 & 21 \\
LLaVA-v1.6-Mistral-7B-HF(M)& \textbf{13.2} & 24.4 & 32.2 \\
GLM-4.1V-9B-Thinking   & 6.2 & 13.4 & 16.2 \\
GLM-4.1V-9B-Thinking(T) & 2.6 & 3.8 & 5.2 \\
GLM-4.1V-9B-Thinking(M) & 12.2 & \textbf{25.6} & \textbf{32.4} \\
\hline
\end{tabular}
\label{tab:grouped_results}
\end{table}
\textbf{Performance Gap and Upper Bound} A significant performance gap exists between the \textbf{Original Query} (lower bound) and the \textbf{Target Query} (oracle rewriting), underscoring the necessity of conversational query rewriting. The fact that even the oracle rewrite falls short of the \textbf{Caption} performance ceiling indicates that the retrieval task itself is challenging and that there remains room for improvement in generating optimal retrieval-oriented queries.

\textbf{Effectiveness of Fine-tuning} Across all models, fine-tuning on our ReCQR dataset yields substantial performance gains compared to the zero-shot capabilities of the base models (e.g., Qwen2.5-VL-7B-Instruct vs. Qwen2.5-VL-7B-Instruct(T)). This demonstrates the value of our dataset in teaching models to resolve conversational context into effective retrieval queries.

\textbf{Text-Only vs. Multimodal Difficulty} All models perform consistently worse on the Multimodal dataset, confirming its significantly greater complexity over Text-Only grounding.

\textbf{Text-Only Dataset} The Text-Only fine-tuned models (T) consistently outperform their multimodal counterparts (M). We suspect that the second stage of multimodal fine-tuning may cause catastrophic forgetting, impairing the text-based dialogue reasoning capabilities acquired during the initial training stage. 

\textbf{Multimodal Dataset} The multimodal training (M) is essential.  These models significantly outperform text-only versions, which lack visual grounding for cross-image references, confirming that visual information is critical for resolving complex multi-image dependencies.  Top performers GLM-4.1V-9B-Thinking and LLaVA-v1.6-Mistral-7B-HF excel on different metrics.

\textbf{Model Comparison} GLM-4.1V-9B-Thinking showed strong performance in the single-image, text-only setting, achieving the highest R@1 score. However, its performance slightly decreased when visual features were added. On the multimodal task, LLaVA-v1.6-Mistral-7B-HF and GLM-4.1V-9B-Thinking demonstrated superior capabilities in leveraging multimodal context for effective rewriting, with each model achieving the top performance on different recall metrics (R@1 for LLaVA and R@5/R@10 for GLM), indicating complementary strengths.
 
\section{Conclusion}
\label{sec:prior}
In this paper, we present ReCQR, a novel dataset for conversational query rewriting in image retrieval spanning both single- and multi-image dialogues.  Experiments with multiple LLMs confirm CQR effectively bridges ambiguous multi-turn queries with retrieval-ready representations, enhancing retrieval performance.


\bibliographystyle{IEEEbib}
\bibliography{strings,main}

\end{document}